\documentclass[aps]{revtex4}
\usepackage{amsmath}
\usepackage{amsfonts}

\begin{document}

\title{Colorless operators in a non-associative quantum theory}
\author{Vladimir Dzhunushaliev\footnote{Senior Associate of the Abdus Salam ICTP}}
\email{dzhun@krsu.edu.kg} 
\affiliation{Dept. Phys. and Microel. Engineer., Kyrgyz-Russian Slavic University, Bishkek, Kievskaya Str. 44, 720021, Kyrgyz Republic}

\date{\today}

\begin{abstract}
The associators/antiassociators for the product of four non-associative operators are  deduced. By analogy with SU(3) gauge theory the notion of colorless (white) operators is 
introduced. Some properties of white operators are considered. It is hypothesized that  white operators do not give any contribution to corresponding associators/antiassociators.  It is suggested that the observables in a non-associative quantum theory correspond to the 
white operators only.
\end{abstract}


\maketitle

\section{Introduction}

Usually in the construction of a new theory one invents a new classical theory and then quantizes it. For example, this was the procedure followed by Yang and Mills in the generalization of classical Maxwell electrodynamics to classical non-Abelian gauge theories. But one can think of another way: one can change the usual quantization prescriptions in order to obtain a new quantum field theory. It is possible that this may give insight into non-perturbative quantum field theory. 
\par
In Ref. \cite{dzhun1} the hypothesis was put forward that the octonionic numbers can be generalized to octonion-like operators. Physically this means that one is dealing with a non-associative quantum theory. Mathematically it means that the corresponding algebra of
operators is generated not only using the ordinary commutations relationships but also by introducing some quantum corrections connected with the rearrangement of brackets in  products of operators.
\par
One of the main problems for the physical interpretation of this hypothesized non-associative quantum theory is the probability interpretation of the theory. For the resolution of this problem we turn to quantum chromodynamics (QCD). In QCD there is the
assumption that free quarks are not observable: one can observe only colorless objects like nucleons and mesons. We will apply this idea to the non-associative quantum theory: \textit{only colorless (white) quantities are observable.} These white objects
correspond to products of two or three non-associative operators. In order that this idea be self-consistent it is necessary that the colorless operators be combined into an
associative subalgebra.

\section{Short history of non-associativity in physics}

In this section we list some general textbooks and monographs on non-associativity, and  
give a short review of modern work into non-associative physics. 
\par 
The monograph \cite{gursey1996} surveys the role of some associative and non-associative 
algebras in particle physics. It concerns the interplay between division algebras -- specifically 
quaternions and octonions. Selected applications of these algebraic structures are 
discussed: quaternion analyticity of Yang-Mills instantons, octonionic aspects of exceptional broken gauge, 
supergravity theories, division algebras in anyonic phenomena and in theories of extended objects in 
critical dimensions. 
\par 
In the book \cite{okubo1995}, the author applies non-associative algebras to physics. The book covers topics 
ranging from algebras of observables in quantum mechanics, to angular momentum and octonions, to division 
algebra, triple-linear products and Yang - Baxter equations. The non-associative gauge theoretic reformulation 
of Einstein's general relativity theory is also discussed.
\par 
A mathematical introduction to non-associative algebras and their application to physics can be found in Ref's \cite{schafer} and \cite{baez}. 
\par 
It has been known since the 1930s that quantum mechanics can be formulated in quaternionic as well as 
complex Hilbert space. The book \cite{adler1995} signals a major conceptual advance and gives a detailed 
development and exposition of quaternionic quantum mechanics for the purpose of determining whether quaternionic
space is the appropriate arena for the long sought-after unification of the standard model forces with 
gravitation. The book also provides an introduction to the problem of formulating quantum field theories in 
quaternionic Hilbert space and concludes with a chapter devoted to discussions on where quaternionic quantum 
mechanics may fit into the physics of unification.
\par 
In Ref. \cite{Manogue:1989ey} the equations of motion arising from the Green-Schwarz Lagrangian for the 
superstring are solved for both commuting and anticommuting variables. The method of solution is to use an 
octonionic formalism for ten-dimensional vectors and spinors, and the solution is given in terms of a 
number of octonion parameters.
\par 
In Ref. \cite{Manogue:1998rv} an octonion formalism is used to give
a new mechanism for reducing 10 spacetime 
dimensions to 4 without compactification. Applying this mechanism to the free, 10-dimensional, 
massless (momentum space) Dirac equation results in a particle spectrum consisting of exactly 3 generations. 
\par 
A new matrix model is described in Ref. \cite{Smolin:2001wc}, based on the exceptional Jordan algebra. 
The author describes a compactification that reproduces, at the one loop level, an octonionic compactification 
of the matrix string theory in which SO(8) is broken to G2. Supersymmetry appears to be related to triality 
of the representations of Spin(8).
\par 
In Refs. \cite{Grossman}, \cite{Jackiw} it is shown that the three-cocycle arises when a representation of a 
transformation group is nonassociative, so that the Jacobi identity fails. The physical meaning is that when 
the translation group in the presence of a magnetic monopole is represented by gauge-invariant operators, a 
(trivial) three-cocycle occurs. The requirement that finite translations be associative leads to Dirac's 
monopole quantization condition. It is interesting that  three-cocycles can be important in the quark model
with U(6)$\times$U(6) algebra. 
\par
Probably the first attempt to apply the octonions in physics was the paper \cite{jordan} describing an octonionic 
quantum mechanics. In Ref.\cite{kugo} it is shown that some relation between octonions and string theory exists. 
In Ref.\cite{merab} an octonionic geometry is described. In this paper a new geometrical interpretation of the 
products of octonionic basis units is presented and the eight real parameters of octonions are interpreted as the 
spacetime coordinates, momentum and energy. In Ref.\cite{dzhun} the non-associativity for the non-perturbative 
quantization of strongly interacting fields was introduced. A generalization of the quantum Hall effect where particles 
move in an eight dimensional space under an $SO(8)$ gauge field is considered in Ref.\cite{Bernevig:2003yz}. The 
underlying mathematics of this particle liquid is that of the last normed division
algebra, the octonions. In Ref.\cite{Nesterov:2000qb}, in the framework of non-associative geometry, a 
unified description of continuum and discrete spacetime is proposed. In Ref.\cite{Nesterov:2004bn} it 
is shown that the non-associative extension of the group U(1) allows one to obtain a consistent theory of point-like 
magnetic monopoles having an arbitrary magnetic charge. Another recent application is the construction of a 
gauge theory on non-associative spaces \cite{Majid:2005kp}-\cite{Benslama:1999ja}. 
\par
In Ref's \cite{Color} \cite{gursey} a possible connection between quark confinement and octonions was investigated. The difference between the approaches presented in Ref. \cite{Color} \cite{gursey} and here is that we consider \textit{octonion-like} operators (not octonions) but in Ref. \cite{Color} \cite{gursey} the quantum theory is based on the octonions not octonion-like operators. In Ref. \cite{gursey} the quark field operators are 
\begin{equation}
  \hat \psi_i = \hat q_i^n u_n 
\label{sec1-10}
\end{equation}
where $u_n$ are the split-octonions and $q_i^{\dagger n}, q_i^n$ are the usual quark creation/annihilation operators. In the presented paper we introduce non-associative operators $Q_i, q_i$ which are the non-associative generalizations of $q_i^{\dagger n}, q_i^n$ operators. 

\section{Some general remarks about non-associative algebra and
associators/antiassociators}

Following the conventions of \cite{dzhun1} we will consider non-associative operators $q_i$ and $Q_i$. These operators are generalizations of the split-octonion numbers $\tilde{q}_i$ and $\tilde{Q}_j$ to  non-associative operators $q_i,Q_i$. One can say that $q_i, Q_i$ and $\tilde{q}_i,\tilde{Q}_j$ are like $q$ and $c$-Dirac numbers in the standard quantum mechanics (according to Dirac's terminology $q-$numbers is noncommuting operators and $c-$numbers is ordinary commuting numbers).
\par 
We now make some remarks about the calculation of the
associators/antiassociators. Let the operator $A_1$ be some product of $n$
non-associative operators $q_i, Q_j$ and $A_2$ be the same operator
but with a different arrangement of the brackets. If $\tilde{q}_i,
\tilde{Q}_j$ are the corresponding split-octonions then $\tilde{A}_1$ and
$\tilde{A}_2$ are the corresponding octonions after the replacement
$q_i \rightarrow \tilde{q}_i$ and $Q_i \rightarrow \tilde{Q}_i$ and
then
\begin{equation}
    \tilde A_2 = \pm \tilde A_1 .
\label{sec0-10}
\end{equation}
This permits us to define the corresponding quantum associator/antiassociator
for the operators
\begin{equation}
    A_2 - \left( \pm A_1 \right) = \mathcal{A}.
\label{sec0-20}
\end{equation}
where $\mathcal{A}$ is the associator for $(+)$ and antiassociator for $(-)$. We will  assume that if $A_{1,2}$ are the product of $n$ operators then the quantum  associator/antiassociator $\mathcal{A}$ either is a new operator or is composed of only $n-1$ operators or less.
\par
In Ref. \cite{dzhun1} the quantum associators/antiassociators of products of three operators are introduced. In this Letter these definitions are a little changed 
\begin{eqnarray}
    \left( Q_mQ_n \right) Q_p + Q_m\left( Q_n Q_p \right) &=&
    \epsilon_{mnp} \mathcal{H}_{3,1} ,
\label{sec1-20a}\\
    \left( q_m q_n \right) q_p - q_m\left( q_n q_p \right) &=&
    \epsilon_{mnp} \mathcal{H}_{3,2} ,
\label{sec1-30}\\
    \left( q_m Q_n \right) q_p + q_m\left( Q_n q_p \right) &=&
    \epsilon_{mnp} \mathcal{H}_{3,3} ,
\label{sec1-40}\\
    \left( q_m q_n \right) Q_p - q_m\left( q_n Q_p \right) &=&
    \epsilon_{mnp} \mathcal{H}_{3,4} ,
\label{sec1-50}\\
    \left( Q_m q_n \right) q_p - Q_m\left( q_n q_p \right) &=&
    \epsilon_{mnp} \mathcal{H}_{3,5} ,
\label{sec1-60}\\
    \left( q_m Q_n \right) Q_p - q_m\left( Q_n Q_p \right) &=&
    \epsilon_{mnp} \mathcal{H}_{3,6} ,
\label{sec1-70}\\
    \left( Q_m q_n \right) Q_p - Q_m\left( q_n Q_p \right) &=&
    \epsilon_{mnp} \mathcal{H}_{3,7} ,
\label{sec1-80}\\
    \left( Q_m Q_n \right) q_p - Q_m\left( Q_n q_p \right) &=&
    \epsilon_{mnp} \mathcal{H}_{3,8}
\label{sec1-90}\\
    q_m\left( Q_n q_n \right) + \left( q_m Q_n \right) q_n &=& q_m ,
\label{sec1-100}\\
    q_n\left( Q_n q_m \right) + \left( q_n Q_n \right) q_m &=& q_m ,
\label{sec1-110}\\
    Q_m\left( q_n Q_n \right) + \left( Q_m q_n \right) Q_n &=& Q_m ,
\label{sec1-120}\\
    Q_n\left( q_n Q_m \right) + \left( Q_n q_n \right) Q_m &=& Q_m .
\label{sec1-130}\\
    \left( q_m Q_m \right) Q_n - q_m\left( Q_m Q_n \right) &=& \rho_1 Q_n ,
\label{sec1-140}\\
   \left( q_m Q_n \right) Q_m - q_m\left( Q_n Q_m \right) &=& \rho_2 Q_n ,
\label{sec1-150}\\
    \left( Q_m q_m \right) q_n - Q_m\left( q_m q_n \right) &=& \gamma_1 q_n ,
\label{sec1-160}\\
    \left( Q_m q_n \right) q_m - Q_m\left( q_n q_m \right) &=& \gamma_2 q_n ,
\label{sec1-170}\\
    \left( Q_m Q_n \right) q_m - Q_m\left( Q_n q_m \right) &=& \delta_1 Q_n ,
\label{sec1-180}\\
    \left( Q_n Q_m \right) q_m - Q_n\left( Q_m q_m \right) &=& \delta_2 Q_n ,
\label{sec1-190}\\
    \left( q_m q_n \right) Q_m - q_m\left( q_n Q_m \right) &=& \tau_1 q_n ,
\label{sec1-200}\\
    \left( q_n q_m \right) Q_m - q_n\left( q_m Q_m \right) &=& \tau_2 q_n .
\label{sec1-210}
\end{eqnarray}
where $\mathcal{H}_{3,i}$ are some operators which can be different for different $i$; $\rho_{1,2}, \delta_{1,2}, \gamma_{1,2}, \tau_{1,2}$ are some numbers and $m,n = 1,2,3$. The self-consistency of the quantum associators and antiassociators \eqref{sec1-20a} - \eqref{sec1-130} has been proved in Ref. \cite{dzhun1} and \eqref{sec1-140} - \eqref{sec1-210} can be proved by similar manner that gives us 
\begin{eqnarray}
    \rho_1 + \rho_2 &=& \delta_1 + \delta_2 ,
\label{sec1-211}\\
    \tau_1 + \tau_2 &=& \gamma_1 + \gamma_2 .
\label{sec1-212}
\end{eqnarray}
The alternativity properties of the octonion-like operators $q_i, Q_i$ remain former. 

\section{The product of two white operators}

Equations \eqref{sec1-100}-\eqref{sec1-210} indicate that if either the left or  right hand sides has products like $q_i Q_i$ then this product does not play a part in the
associator/antiassociator. One can see that there is some similarity between 
$q_i, Q_j$ and the annihilation and creation operators of quarks. The indices $i=1,2,3$ correspond to SU(3) colors =~$red, green, blue$. Because of this we call these products $q_i Q_j$ \textit{colorless(white)} operators, i.e. the white operators, $\mu_i,
i=1,2,3$ are
\begin{equation}
    \mu_i = q_i Q_i 
\label{sec1-270}
\end{equation}
here $i$ is not summed over. Comparing with QCD $q_i$ and $Q_i (i=1,2,3)$ are the  generalizations of the annihilation and creation operators of quarks correspondingly. This 
comparison indicates that there are a few white operators with three factors
\begin{eqnarray}
    n_1 &=& \left ( q_1 q_2 \right ) q_3
\label{sec1-280}\\
    n_2 = n_1^\dag &=&  Q_3 \left ( Q_2 Q_1 \right ),
\label{sec1-290}
\end{eqnarray}
where these operators are Hermitian conjugates and any permutations
of indices are possible.
\par
The quantum associators/antiassociators in \eqref{sec1-100}-\eqref{sec1-210} are deduced in such a way that for any permutation of two factors in 
\eqref{sec1-100}-\eqref{sec1-210} the result would not depend on the way in which the 
brackets are arranged \cite{dzhun1}. Additionally we will demand the same for the product of two  colorless operators that is considered in this section.
\par
For the calculation of the product of two white operators $(q_iQ_i)$
and $(q_j Q_j)$ we will introduce the following non-associative rules
\begin{eqnarray}
  \left( q_i Q_i \right)\left( q_j Q_j \right) &=&
  \left( \left( q_i Q_i \right) q_j \right) Q_j +
  \alpha_1 \left( q_j Q_j \right)
  + \beta_1
\label{sec3-10}\\
  \left( q_i \left( q_j Q_j \right) \right) Q_i &=&
  q_i \left( \left( q_j Q_j \right) Q_i \right) +
  \alpha'_1 \left( q_i Q_i \right) +
  \beta'_1 ,
\label{sec3-20}\\
  \left( q_i Q_i \right)\left( q_j Q_j \right) &=& q_i \left( Q_i
\left( q_j Q_j \right)\right) + \alpha_2 \left( q_i Q_i \right) +
\beta_2
\label{sec3-30}
\end{eqnarray}
here $i,j$ are not summed over. Now we consider the following product
\begin{equation}
\begin{split}
    \left( q_1 Q_1 \right)\left( q_2 Q_2 \right) =
    &\left( \left( q_1 Q_1 \right) q_2 \right)Q_2 +
    \alpha_1 \left( q_2 Q_2 \right)
    + \beta_1 =
    - \left( q_1 \left( Q_1q_2 \right) \right) Q_2 + q_2 Q_2 +
    \alpha_1 \left( q_2 Q_2 \right)
    + \beta_1 = \\
    &\left( q_1 \left( q_2 Q_1 \right) \right) Q_2 + q_2 Q_2 +
    \alpha_1 \left( q_2 Q_2 \right)
    + \beta_1 =
    \left( \left( q_1q_2 \right) Q_1 \right) Q_2 +
    \alpha_1 \left( q_2 Q_2 \right)
    + \beta_1 = \\
    &- \left( \left( q_2 q_1 \right) Q_1 \right) Q_2 +
    \alpha_1 \left( q_2 Q_2 \right)
    + \beta_1 =
    - \left( q_2 \left( q_1 Q_1 \right) \right) Q_2 - q_2 Q_2 +
    \alpha_1 \left( q_2 Q_2 \right) +
    \beta_1 = \\
    & - q_2 \left( \left( q_1 Q_1 \right) Q_2 \right)- q_2 Q_2 +
    \left( q_2 Q_2 \right) \left( \alpha_1 - \alpha'_1 \right) +
    \left( \beta_1 - \beta'_1 \right) = \\
    &-q_2 \left( q_1 \left( Q_1 Q_2 \right) \right) +
    \left( q_2 Q_2 \right) \left( \alpha_1 - \alpha'_1 \right) +
    \left( \beta_1 - \beta'_1 \right) = \\
    &q_2 \left( q_1 \left( Q_2 Q_1 \right) \right) +
    \left( q_2 Q_2 \right) \left( \alpha_1 - \alpha'_1 \right) +
    \left( \beta_1 - \beta'_1 \right) = \\
    & q_2 \left( \left( q_1 Q_2 \right) Q_1 \right) + q_2 Q_2 +
    \left( q_2 Q_2 \right) \left( \alpha_1 - \alpha'_1 \right) +
    \left( \beta_1 - \beta'_1 \right) = \\
    &- q_2 \left( \left( Q_2 q_1 \right) Q_1 \right) + q_2 Q_2 +
    \left( q_2 Q_2 \right) \left( \alpha_1 - \alpha'_1 \right) +
    \left( \beta_1 - \beta'_1 \right) = \\
    & q_2 \left( Q_2 \left( q_1 Q_1 \right) \right) +
    \left( q_2 Q_2 \right) \left( \alpha_1 - \alpha'_1 \right) +
    \left( \beta_1 - \beta'_1 \right) = \\
    \left( q_2 Q_2 \right)\left( q_1 Q_1 \right) + &
    \left( q_2 Q_2 \right) \left( \alpha_1 - \alpha'_1 - \alpha_2 \right) +
    \left( \beta_1 - \beta'_1 - \beta_2 \right) .
\label{sec3-50}
\end{split}
\end{equation}
This is the first way. Another way is
\begin{equation}
\begin{split}
    \left( q_1 Q_1 \right)\left( q_2 Q_2 \right) =
    &q_1 \left( Q_1 \left( q_2 Q_2 \right)\right) +
    \alpha_2 \left( q_1 Q_1 \right)
    + \beta_2 =
    -q_1 \left(\left( Q_1 q_2 \right) Q_2 \right) + q_1 Q_1 +
    \alpha_1 \left( q_2 Q_2 \right)
    + \beta_1 = \\
    &q_1 \left(\left( q_2 Q_1 \right) Q_2 \right) + q_1 Q_1 +
    \alpha_1 \left( q_2 Q_2 \right)
    + \beta_1 =
    q_1 \left( q_2 \left( Q_1 Q_2 \right) \right) +
    \alpha_1 \left( q_2 Q_2 \right)
    + \beta_1 = \\
    &- q_1 \left( q_2 \left( Q_2 Q_1 \right)\right) +
    \alpha_1 \left( q_2 Q_2 \right)
    + \beta_1 =
    -q_1 \left(\left( q_2 Q_2 \right) Q_1 \right) - q_1 Q_1 +
    \alpha_1 \left( q_2 Q_2 \right)
    + \beta_1 = \\
    & - \left( q_1 \left( q_2 Q_2 \right)\right) Q_1 - q_1 Q_1 +
    \left( q_1 Q_1 \right) \left( \alpha_2 + \alpha'_1 \right) +
    \left( \beta_2 + \beta'_2 \right) = \\
    &- \left(\left( q_1 q_2 \right) Q_2 \right) Q_1 +
    \left( q_1 Q_1 \right) \left( \alpha_2 + \alpha'_1 \right) +
    \left( \beta_2 + \beta'_1 \right) = \\
    &\left(\left( q_2 q_1 \right) Q_2 \right) Q_1 +
    \left( q_1 Q_1 \right) \left( \alpha_2 + \alpha'_1 \right) +
    \left( \beta_2 + \beta'_1 \right) = \\
    & \left( q_2 \left( q_1 Q_2 \right)\right) Q_1 + q_1 Q_1 +
    \left( q_1 Q_1 \right) \left( \alpha_2 + \alpha'_1 \right) +
    \left( \beta_2 + \beta'_1 \right) = \\
    &- \left( q_2 \left( Q_2 q_1 \right)\right) Q_1 + q_1 Q_1 +
    \left( q_1 Q_1 \right) \left( \alpha_2 + \alpha'_1 \right) +
    \left( \beta_2 + \beta'_1 \right) = \\
    &\left(\left( q_2 Q_2 \right) q_1 \right) Q_1 +
    \left( q_1 Q_1 \right) \left( \alpha_2 + \alpha'_1 \right) +
    \left( \beta_2 + \beta'_1 \right) = \\
    \left( q_2 Q_2 \right)\left( q_1 Q_1 \right) +
    & \left( q_1 Q_1 \right) \left( \alpha_2 + \alpha'_1 - \alpha_1 \right) +
    \left( \beta_2 + \beta'_1 - \beta_1 \right) .
\label{sec3-60}
\end{split}
\end{equation}
Comparing equations \eqref{sec3-50} and \eqref{sec3-60} we see that
one should have
\begin{eqnarray}
    \alpha_1 - \alpha'_1 - \alpha_2 &=& \alpha_2 + \alpha'_1 -
    \alpha_1 = 0,
\label{sec3-70}\\
    \beta_1 - \beta'_1 - \beta_2 &=& \beta_2 + \beta'_1 - \beta_1 =
    \beta .
\label{sec3-80}
\end{eqnarray}
Then we can calculate the commutator
\begin{equation}
    \left( q_1 Q_1 \right)\left( q_2 Q_2 \right) -
    \left( q_2 Q_2 \right)\left( q_1 Q_1 \right) = \beta .
\label{sec3-90}
\end{equation}
The pair of indices 1 and 2 can be replaced by any other.

\section{Invisibility principle}
In \eqref{sec3-50} and \eqref{sec3-60} we have used an invisibility principle. 
This principle tells us that \textit{in a calculation with rearrangements of 
brackets in a product the white operator is invisible. This implies that if 
there is a colorless operator on the lhs or rhs then the corresponding quantum 
associator/antiassociator does not have any operators which 
were in the colorless operator.}
\par
For example, for the product of three operators we have
\begin{equation}
    q_m \left( Q_n q_n \right) + \left( q_m Q_n \right) q_n = q_m
\label{sec2-10b}
\end{equation}
here the white operator is $Q_n q_n$ and the antiassociator has 
only $q_m$. This explains why in Eqs. \eqref{sec1-100} - \eqref{sec1-210} 
the quantum associators (antiassociators) have only one operator $q$ or $Q$. 
\par 
For the product of four operators \eqref{sec3-10} we had
\begin{equation}
  \left( q_i Q_i \right)\left( q_j Q_j \right) -
  \left( \left( q_i Q_i \right) q_j \right) Q_j =
  \alpha_1 \left( q_j Q_j \right)
  + \beta_1
\label{sec2-20}
\end{equation}
also here $q_i Q_i$ is the white operator and consequently the
associator can have only a term like $q_j Q_j$. According to the
invisibility principle we have two white operators in equation
\eqref{sec2-20}, $\left( q_i Q_i \right)$ and $\left( q_j Q_j
\right)$, and consequently
\begin{equation}
  \alpha_1 = \alpha_2 = 0.
\label{sec2-25}
\end{equation}
Then from equation \eqref{sec3-70} we see that
\begin{equation}
  \alpha'_1 = 0
\label{sec2-27}
\end{equation}
and consequently
\begin{eqnarray}
  \left( q_i Q_i \right)\left( q_j Q_j \right) &=&
  \left( \left( q_i Q_i \right) q_j \right) Q_j + \beta_1
\label{sec2-28a}\\
  \left( q_i \left( q_j Q_j \right) \right) Q_i &=&
  q_i \left( \left( q_j Q_j \right) Q_i \right) + \beta'_1 ,
\label{sec2-28b}\\
  \left( q_i Q_i \right)\left( q_j Q_j \right) &=&
  q_i \left( Q_i \left( q_j Q_j \right)\right) + \beta_2 ,
\label{sec2-28c}
\end{eqnarray}
Let us calculate the Hermitian conjugation of equation
\eqref{sec2-28b}
\begin{equation}
    \left( q_i \left( q_j Q_j \right) \right) Q_i =
  q_i \left( \left( q_j Q_j \right) Q_i \right) - {\beta'_1}^*
\label{sec2-29}
\end{equation}
comparing this with \eqref{sec2-28b} we see that
\begin{equation}
    \beta'_1 = - {\beta'_1}^* .
\label{sec2-29a}
\end{equation}
This means that $\beta'_1$ is a pure imaginary number
\begin{equation}
    \beta'_1 = - {\beta'_1}^* = i \hbar_2
\label{sec2-29b}
\end{equation}
where $\hbar_2$ is some constant. The Hermitian conjugation of
\eqref{sec2-28a} gives us
\begin{equation}
  \left( q_i Q_i \right)\left( q_j Q_j \right) =
  q_i \left( Q_i \left( q_j Q_j \right)\right) + \beta^*_1
\label{sec2-29c}
\end{equation}
here we have replaced $i \leftrightarrow j$. Comparing with
\eqref{sec2-28c} we have
\begin{equation}
  \beta_2 = \beta^*_1 .
\label{sec2-29d}
\end{equation}
For simplicity we suppose that $\beta_{1,2}$ are imaginary
numbers, too
\begin{equation}
  \beta_2 = \beta^*_1 = - \frac{i}{2} \hbar_2
\label{sec2-29e}
\end{equation}
where $\hbar'_2$ is some constant.
In this case \eqref{sec2-28a}-\eqref{sec2-28c} are
\begin{eqnarray}
  \left( q_i Q_i \right)\left( q_j Q_j \right) &=&
  \left( \left( q_i Q_i \right) q_j \right) Q_j + \frac{i}{2} \hbar_2
\label{sec2-29f}\\
  \left( q_i \left( q_j Q_j \right) \right) Q_i &=&
  q_i \left( \left( q_j Q_j \right) Q_i \right) + i \hbar_2 ,
\label{sec2-29g}\\
  \left( q_i Q_i \right)\left( q_j Q_j \right) &=&
  q_i \left( Q_i \left( q_j Q_j \right)\right) - \frac{i}{2} \hbar_2 ,
\label{sec2-29h}
\end{eqnarray}
Finally the commutator \eqref{sec3-90} is
\begin{equation}
    \left( q_1 Q_1 \right)\left( q_2 Q_2 \right) -
    \left( q_2 Q_2 \right)\left( q_1 Q_1 \right) = 0
\label{sec2-29i}
\end{equation}
The application of this principle to the product of three white
operators gives us
\begin{equation}
    \biggl[
        \left( q_i Q_i \right) \left( q_j Q_j \right)
    \biggl]
    \left( q_k Q_k \right) =
    \left( q_i Q_i \right)
    \biggl[
        \left( q_j Q_j \right) \left( q_k Q_k \right)
    \biggl]
\label{sec2-30}
\end{equation}
here the repeated indices are not summed over. Equation \eqref{sec2-30} tells us that 
the colorless operators are associative at least at the level of the product of three white operators.

\section{Observables in a non-associative quantum theory}

The next important question in a non-associative quantum theory is the probability interpretation. Generally speaking, the non-associative theory has problems here because any non-associative quantum theory may have a matrix representation but such a representation will be non-associative. We want to avoid this problem using the experience from QCD. In QCD free quarks are unobservable. Only colorless mesons and nucleons are observable. We would like to apply this lesson from QCD to non-associative quantum theory: \textit{in a non-associative quantum theory the physical observables are represented only by white operators.} In this case we avoid the problems of non-associative quantum theory since the colorless operators are associative operators according to the invisibility principle.

\section{Discussion and conclusions}

In this paper we have obtained some quantum associators/antiassociators for the product of four operators where we considered white operators  only. The corresponding quantum associators/antiassociators were deduced using the invisibility principle which implies that white operators are invisible for the determination of the corresponding associators/antiassociators.
\par
To resolve the problem connected with the probability interpretation of a non-associative 
quantum theory we postulated that the observables in this theory are represented only by
colorless (white) operators. This means that the real observables are composite objects with an inner structure which is unobservable. The unobservable degrees of freedom are similar to ``hidden variables'' in the corresponding ``hidden variables theory'' which is alternative to standard quantum mechanics. The difference is that in such theory the hidden parameters were introduced as classical degrees of freedom, but in the non-associative quantum theory the hidden parameters are quantum degrees of freedom.
\par
Let us note two alternative views of these invisible degrees of freedom: the first is that the non-associativity is a fundamental property; the second is that the non-associativity is some approximate description of a strongly interacting quantum system (as in QCD or gravity). In the first case some aspects of Nature will always be hidden. In the second case the white object is similar to a nucleon where three quarks are confined by the SU(3) gauge field. In this case the non-associative quantum theory is  similar to the four-fermion model of the weak interaction before the introduction of $SU(2) \times U(1)$ Standard Model.
\par
Another problem is how one can define field operators from the creation/annihilation $q_i, Q_i$ operators considered here ? What equations describe the dynamics of such a
non-associative quantum field ? In this connection we note that in Ref. \cite{Gogberashvili:2005cp} an octonionic version of Dirac's equation was formulated.
\par
The idea that the real elementary particles are composite fields which are constructed from more fundamental constituents is not new. In Ref. \cite{Bars:1980ub} a model is constructed where ``ternons'' are the fundamental constituents. In Ref. \cite{Pati:1980rx} a preonic, composite model of quarks and leptons is presented. The difference with the non-associative approach presented here is that in our case \textit{the constituent non-associative physical
quantities are unobservable in principle.} But the question as to whether non-associativity plays a role in Nature is far from being resolved since the model presented here is a simple, toy model. Our aim was only to show that it is possible to construct a quantum theory with (anti)associative relationships in addition to the ordinary (anti)commutations relationships. The real algebra of strongly interacting quantum fields may be much more complicated.

\section*{Acknowledgment}

I am grateful to the ICTP for financial support and the invitation to research and the referee for offering the term ``quantum associator'' and D. Singleton for the fruitful remarks.


\begin{thebibliography}{99}

\bibitem{dzhun1}
V.~Dzhunushaliev, ``A non-associative quantum mechanics'', to be published in Found. Phys. Lett., hep-th/0502216. 

\bibitem{gursey1996}
Feza G\"ursey and Chia-Hsiung Tze, \textit{On the Role of Division, Jordan and Related Algebras in Particle Physics}, World Scientific, Singapore, 1996.

\bibitem{okubo1995}
Susumo Okubo, \textit{Introduction to Octonion and Other Non-Associative Algebras in Physics}, Cambrodge University Press, Cambridge, 1995.

\bibitem{schafer}
R. Schafer,
\textit{Introduction to Non-Associative Algebras}
(Dover, New York, 1995);\\
T. A. Springer and F. D. Veldkamp, \textit{Octonions, Jordan
Algebras and Exceptional Groups},
Springer Monographs in Mathematics (Springer, Berlin, 2000);\\
J. L$\tilde{o}$hmus, E. Paal and L. Sorgsepp, Acta Appl. Math.,
\textbf{50}, 3 (1998).

\bibitem{baez}
J.~C.~Baez,
``The Octonions,'' 
Bull. Amer. Math. Soc., \textbf{39}, 145-205 (2002), 
math.ra/0105155.

\bibitem{adler1995}
Stephen L. Adler, \textit{Quaternionic Quantum Mechanics and Quantum Fields}, Oxford University Press, New York, 1995. 
 
\bibitem{Manogue:1989ey}
  C.~A.~Manogue and A.~Sudbery,
  Phys.\ Rev.\ D {\bf 40}, 4073 (1989).

\bibitem{Manogue:1998rv}
  C.~A.~Manogue and T.~Dray,
  Mod.\ Phys.\ Lett.\ A {\bf 14}, 99 (1999)
  [arXiv:hep-th/9807044].
  
\bibitem{Smolin:2001wc}
  L.~Smolin,
  arXiv:hep-th/0104050.

\bibitem{Grossman}
B. Grossman,
``A Three Cocycle in Quantum Mechanics,''
\textit{Phys. Lett.} \textbf{B152}, 
93, (1985).

\bibitem{Jackiw}
R. Jackiw,
``3 - Cocycle in Mathematics and Physics,''
\textit{Phys. Rev. Lett.} \textbf{54},
159, (1985).

\bibitem{jordan}
Pascual Jordan, John von Neumann, Eugene Wigner,
``On an algebraic generalization of the quantum mechanical formalism,''
\textit{Ann.\ Math.} \textbf{35}, 29-64 (1934).

\bibitem{kugo}
T.\ Kugo and P.--K.\ Townsend,
``Supersymmetry and the division algebras,''
\textit{Nucl.\ Phys.} \textbf{B221}, 357-380 (1983).

\bibitem{merab}
M. Gogberashvili,
``Octonionic Geometry,''
hep-th/0409173.

\bibitem{dzhun}
V. Dzhunushaliev,
``Non-perturbative operator quantization of strongly interacting fields,''
\textit{Found. Phys. Lett.} \textbf{16}, 57 (2003).

\bibitem{Bernevig:2003yz}
B.~A.~Bernevig, J.~p.~Hu, N.~Toumbas and S.~C.~Zhang,
``The Eight Dimensional Quantum Hall Effect and the Octonions,''
\textit{Phys.\ Rev.\ Lett.}  \textbf{91}, 236803 (2003).

\bibitem{Nesterov:2000qb}
A.~I.~Nesterov and L.~V.~Sabinin,
``Nonassociative geometry: Towards discrete structure of spacetime,''
\textit{Phys.\ Rev.} \textbf{D62}, 081501 (2000).

\bibitem{Nesterov:2004bn}
A.~I.~Nesterov,
``Three-cocycles, nonassociative gauge transformations and Dirac's monopole,''
\textit{Phys.\ Lett.} \textbf{A328}, 110 (2004).

\bibitem{Majid:2005kp}
S.~Majid, ``Gauge theory on nonassociative spaces,''
math.qa/0506453.

\bibitem{deMedeiros:2004wb}
P.~de Medeiros and S.~Ramgoolam,
JHEP, \textbf{0503}, 072 (2005).

\bibitem{Ramgoolam:2003cs}
S.~Ramgoolam,
JHEP, \textbf{0403}, 034 (2004).

\bibitem{Benslama:1999ja}
A.~Benslama and N.~Mebarki,
JHEP, \textbf{0002}, 018 (2000).

\bibitem{Color} 
M. G\"unaydin and F. G\"ursey
Lett. Nuovo Cimento, \textbf{6} 401, (1973); \\
J. Math. Phys., \textbf{14}, 1651 (1973); \\
K. Morita, Prog. Theor. Phys., \textbf{65} 787, (1981).

\bibitem{gursey} 
M. G\"unaydin and F. G\"ursey
Phys. Rev., \textbf{D9}, 3387 (1974).

\bibitem{Gogberashvili:2005cp}
M.~Gogberashvili, ``Octonionic version of Dirac equations,''
hep-th/0505101.

\bibitem{Bars:1980ub}
I.~Bars and M.~Gunaydin,
Phys.\ Rev.\ D {\bf 22}, 1403 (1980).

\bibitem{Pati:1980rx}
J.~C.~Pati, A.~Salam and J.~Strathdee,
Nucl.\ Phys.\ B {\bf 185}, 416 (1981).

\end{thebibliography}
\end{document}